\documentclass{article}
\usepackage{spconf,amsmath,graphicx}
\usepackage{subcaption}
\usepackage{amsfonts}
\usepackage{url}

\usepackage{multirow}
\usepackage{etoolbox}
\makeatletter
\let\@titlehook=\relax
\apptocmd{\@maketitle}{\@titlehook}{}{}
\newcommand{\titlehook}[1]{\def\@titlehook{#1}}
\makeatother

\title{MetaH2: A Snapshot Metasurface HDR Hyperspectral Camera}
\name{Yuxuan Liu and Qi Guo\thanks{Copyright 2025 IEEE. Published in 2025 IEEE International Conference on Image Processing (ICIP), scheduled for 14-17 September 2025 in Anchorage, Alaska, USA. Personal use of this material is permitted. However, permission to reprint/republish this material for advertising or promotional purposes or for creating new collective works for resale or redistribution to servers or lists, or to reuse any copyrighted component of this work in other works, must be obtained from the IEEE. Contact: Manager, Copyrights and Permissions / IEEE Service Center / 445 Hoes Lane / P.O. Box 1331 / Piscataway, NJ 08855-1331, USA. Telephone: + Intl. 908-562-3966.}}
\address{Elmore Family School of Electrical and Computer Engineering \\
Purdue University, West Lafayette, IN, 47907, USA \\
{\small\url{{liu3910, qiguo}@purdue.edu}}
}

\begin{document}
%
%

\titlehook{\begin{center}
    \includegraphics[width=\textwidth]{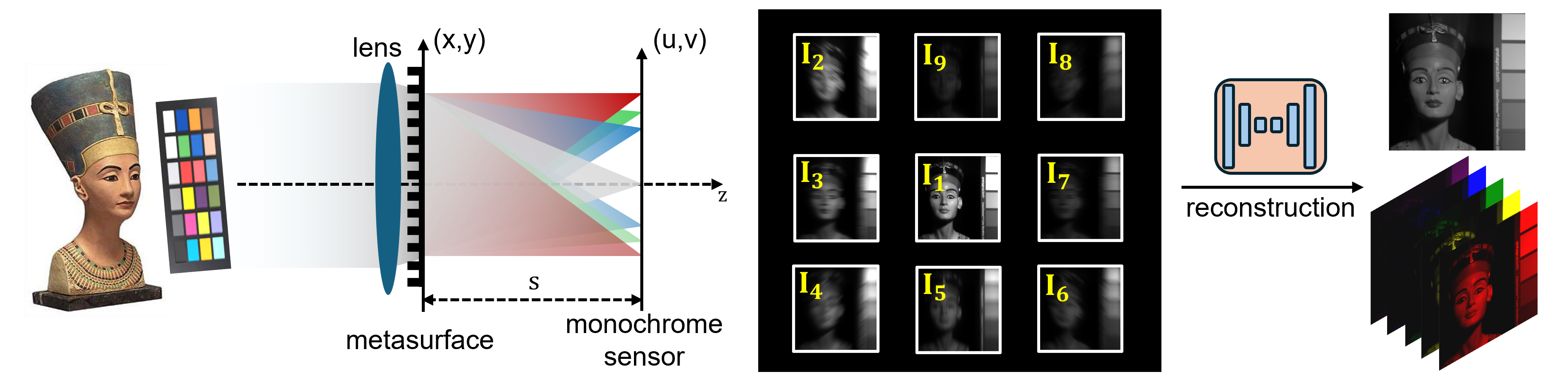}
    \captionof{figure}{Principle. MetaH2 uses an optical system composed of a lens and a custom-engineered metasurface that splits and focuses incoming light onto multiple regions of a monochrome sensor. Each focused spot has a distinct energy and dispersion. Consequently, the system can capture multiple sub-images $I_{1-9}$ of the scene—each with different dynamic ranges and chromatic aberrations—in a single shot. These sub-images are then processed by a reconstruction algorithm to generate both a high-dynamic-range (HDR) image and a hyperspectral datacube simultaneously.  }
    \label{fig:teaser}
    \end{center}}

\maketitle

\begin{abstract}
We present a metasurface camera that jointly performs high-dynamic range (HDR) and hyperspectral imaging in a snapshot. The system integrates exposure bracketing and computed tomography imaging spectrometry (CTIS) by simultaneously forming multiple spatially multiplexed projections with unique power ratios and chromatic aberrations on a photosensor. The measurements are subsequently processed through a deep reconstruction model to generate an HDR image and a hyperspectral datacube. Our simulation studies show that the proposed system achieves higher reconstruction accuracy than previous snapshot hyperspectral imaging methods on benchmark datasets. We assemble a working prototype and demonstrate snapshot reconstruction of 60 dB dynamic range and 10 nm spectral resolution from 600 nm to 700 nm on real-world scenes from a monochrome photosensor.
\end{abstract}
\section{Introduction}
The high-dynamic range (HDR) and hyperspectral imaging can record an object's full brightness or spectral profiles, providing critical insights into its intrinsic properties, including material composition, shape curvature, and chemical signature. Traditionally, either approach requires recording multiple sequential measurements with varying optical configurations~\cite{debevec2023recovering,mouroulis2000design}, which introduces challenges for dynamic scenes, where misalignments between captures necessitate complicated post-processing algorithms. People have proposed approaches that can perform either reconstruction from a single shot of measurement~\cite{metzler2020deep,baek2017compact}. However, \textit{jointly} reconstructing both HDR and hyperspectral (H2) information in a snapshot has been less studied to our knowledge. 

We propose MetaH2, a computational camera that reconstructs H2 in a single shot using a metasurface in its optics. A metasurface is a planar glass substrate patterned with nanoscale structures to manipulate the traversing light. It has recently gained attention in the computational imaging community for its compactness and multifunctionality in imaging systems. MetaH2 utilizes a specially designed metasurface that splits an incident light beam into multiple directions, each with a distinct portion of energy and dispersion direction. The image formation process jointly achieves exposure bracketing 
and computed tomography imaging spectrometry (CTIS)~\cite{okamoto1991simultaneous}, resulting in the simultaneous formation of images with different exposure values and chromatic fringing. These measurements are then processed by a deep neural network architecture to reconstruct an HDR image and the hyperspectral datacube jointly. 

The chromatic aberration of metasurfaces has been a constant obstacle to applying this technology to imaging~\cite{he2022optical}, often requiring narrow bandpass filters to limit incident light to a single wavelength.
In contrast, this work leverages the metasurface's inherent chromaticity as an optical encoder for the hyperspectral information, which broadens the operational bandwidth, significantly improving the light efficiency compared to previous metasurface imagers~\cite{brookshire2024metahdr}.

We conduct both simulation and real-world experiments to study the proposed approach. First, we compare MetaH2 with other snapshot hyperspectral imaging solutions on synthetic data, where MetaH2 demonstrates the highest reconstruction accuracy. 
Then, we build a prototype demonstrating the H2 reconstruction on real-world scenes. The prototype achieves a 60 dB dynamic range and 10 nm spectral resolution for a working bandwidth from 600 nm to 700 nm. 

The contributions of our work include:
\begin{itemize}
    \item A novel metasurface imaging modality for simultaneous exposure bracketing and CTIS, 
    \item A comprehensive analysis of MetaH2 compared to previous snapshot hyperspectral imaging solutions, and 
    \item A working prototype, MetaH2, that performs joint HDR and hyperspectral imaging.
\end{itemize}
\section{Related Work}
\textbf{Snapshot HDR imaging.} Past efforts to snapshot HDR imaging have mostly focused on engineering the photosensors. People have demonstrated photodetectors with mosaicked optical density~\cite{nayar2000high}, exposure~\cite{cho2014single}, or polarization~\cite{wu2020hdr}. People have also exploited novel photon transducers, including single-photon avalanche diode~\cite{ogi2021124}, modulo camera~\cite{so2022mantissacam}, and quanta image sensors~\cite{gnanasambandam2020hdr}. Beyond these, optical solutions have also been investigated. Brookshire et al.~\cite{brookshire2024metahdr} design a multifunctional metasurface that simultaneously forms nine exposure-bracketed images, but it only operates on a 10~nm bandwidth of the incident light to mitigate the metasurface's chromatic aberration. To increase the light efficiency of these metasurface HDR cameras, there is a fundamental need to increase their working bandwidth.\\

\noindent\textbf{Snapshot hyperspectral imaging.} Classic hyperspectral measurement relies on spatial or temporal scanning to capture every 2D slice of the hyperspectral data cube, causing long acquisition time~\cite{gat2000imaging}.
Snapshot hyperspectral imaging targets this issue. Coded aperture snapshot spectral imaging (CASSI) harness the principles of compressive sensing by replacing the slit of a traditional spectrometer with a broader field stop, inside which is inserted a binary-coded mask~\cite{wagadarikar2008single}. However, the accommodation of the mask requires the use of imaging and relay lenses, which significantly complicates the hardware design. There is also research on designing diffractive optical elements with spectrally varying point spread functions~\cite{baek2021single,liu2022miniaturized}.
Despite their compact configurations, these designs only report effective imaging performance within a constrained dynamic range. 
A CTIS imager uses a grating to form multiple measurements with different dispersions simultaneously. However, building gratings with accurately controlled brightness differences between orders of diffraction patterns has been reported to be challenging~\cite{habel2012practical}. Our work builds upon the principles of CTIS while addressing the inherent shortcomings of traditional grating-based methods, leveraging metasurfaces to generate differently dispersed images with accurate energy control.\\

\noindent \textbf{Metasurface-based computational imaging.} Compared to traditional refractive optics, metasurfaces offer great versatility in light modulation and enhanced miniaturization at the cost of strong chromaticity. Thus, most existing metasurface imagers and visual sensors have been reported to operate on a narrow bandwidth
~\cite{yang2023monocular,shen2023monocular}. Recently, people have demonstrated metasurface imagers working with a broader bandwidth of incident light, e.g., $>$100~nm, by utilizing jointly designed postprocessing computation to mitigate the chromatic aberration in the measurement~\cite{hu2016plasmonic,phan2019high} or to leverage such chromaticity as a source to reconstruct more scene information from the measurement~\cite{shi2024learned,hazineh2024grayscale}. 

\section{Image Formation}
\subsection{Optical Design}
\label{secsec:optical}
Our optical design is illustrated in Fig.~\ref{fig:teaser}. It uses an achromatic refractive lens paired with a polarization-insensitive $V$-beamsplitting metasurface as the optical assembly. The working bandwidth of the system is $[\lambda_{\text{low}}, \lambda_{\text{high}}]$. Its point spread function (PSF) at wavelength $\lambda$ is:
\begin{align}
    h(u,v,\lambda) = \sum_{i=1}^V \alpha_i h(u - u_i(\lambda), v - v_i(\lambda)),
    \label{eq:optical-model-full}
\end{align}
where $h(u, v)$ is the PSF of the optical system with only the achromatic refractive lens. The PSF $h(u,v,\lambda)$ contains $V$ repetition of $h^*(u, v)$ centered at $(u_i(\lambda), v_i(\lambda))$, each weighted by the power ratio $\alpha_i$.\\

\begin{figure}[h]
    \centering
    \includegraphics[width=1\linewidth]{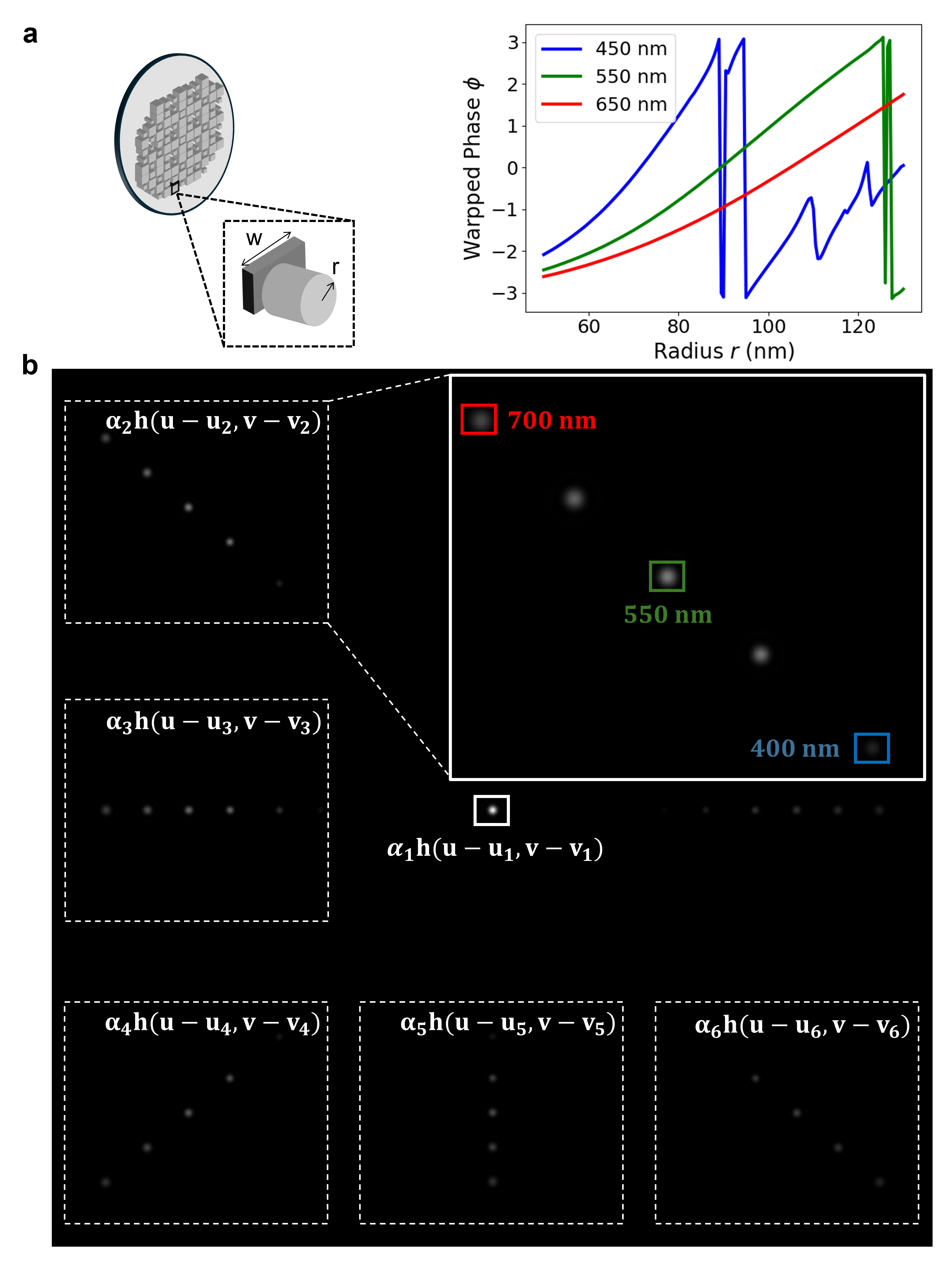}
    \caption{Optics design. (a) We choose the nano-cylinder as the building block on each nanocell so that its phase profile is parametrized by the radius. (b) Simulated PSFs across the full visible band using~\cite{hazineh2022d}. For the index $i=1$, the metasurface has a constant phase profile, producing a PSF that matches the shape of $h(u,v)$. The zoomed-in figure shows the PSF generated by $\phi_{ms,2}(x,y,\lambda)$. In the Fresnel region, the metasurface produces PSFs shifting the locations but remaining their shapes at each wavelength with variations in brightness representing energy differences.}
    \label{fig:simulated_psf}
\end{figure}

\noindent \textbf{Metasurface Design.} The beamsplitting metasurface is designed by interleaving $V$ deflective metasurfaces. For each deflective metasurface, we set its phase delay at the central wavelength $\lambda_c = \frac{\lambda_{\text{low}}+ \lambda_{\text{high}}}{2}$ to be:
\begin{align}
    \phi_{\text{ms},i}(x,y,\lambda_c)=\frac{2\pi}{\lambda_c}(\alpha_i x+\beta_i y), i=1,\cdots, V.
    \label{eq:ms_phase}
\end{align}
where $\alpha_{i}=\frac{u_i(\lambda_c)}{\sqrt{u_i(\lambda_c)^2+v_i(\lambda_c)^2+s^2}}$ and $\beta_i = \frac{v_i(\lambda_c)}{\sqrt{u_i(\lambda_c)^2+v_i(\lambda_c)^2+s^2}}$ are direction cosines of the outgoing wave. Given this phase delay profile, the nanostructure design process employs a standardized, cell-based approach in which the metasurface is represented as a two-dimensional array of uniform nanocells arranged on a regular grid. Under this modeling, the phase delay profile of the metasurface is spatially discretized as $\phi_{\text{ms},i}(p,q,\lambda)$, where $(p,q)\in \mathbb{Z}^2$ denotes the nanocell indices, corresponding to spatial coordinates $x=pw$ and $y=qw$. Here, $w$ is the pitch of each cell. Similar to Brookshire et al.~\cite{brookshire2024metahdr}, we select nano-cylinders with a fixed height as the basic building block on the nanocell. See Fig.~\ref{fig:simulated_psf}a. This choice simplifies the design by associating the desired phase delay $\phi$ with the cylinder radius of a nanocell $r$ (Fig.~\ref{fig:simulated_psf}a). The underlying relationship can be pre-computed using an FDTD simulator and stored in a library, which contains the phase delay as a function of sampled radii and operating wavelengths: $(r, \lambda) \rightarrow \phi$. We determine the nanocell radius of the $i$th deflective metasurface, $r_i(p,q)$, by finding the nano-cylinder radius with the closest phase delay to the designed one at the central wavelength.

The shape parameters $r_i(p,q)$ for individual deflective metasurfaces can be utilized to design the overall shape profile $r(p,q)$ by randomly interleaving the metasurfaces according to \cite{brookshire2024metahdr}. This approach is expressed as:
\begin{align}
\begin{split}
    r(p,q)&=r_i(p,q)\\
    i\sim \text{Multinomial}&\left(\frac{\sqrt{\alpha_1}}{\sum\sqrt{\alpha_i}},\dots,\frac{\sqrt{\alpha_V}}{\sum\sqrt{\alpha_i}}\right)  
\end{split}      
\end{align}
where $\alpha_i=(\frac{1}{2})^i$ specifies the designed power ratio of multiplexes, with each successive index halving the power. By combining multiple metasurfaces in this probabilistic manner, we design a $V$-beamsplitting metasurface that mimics the grating functionality of a conventional CTIS system, forming multiple measurements with different dispersion directions and, furthermore, enabling bracketed exposure. 

As shown in Fig.~\ref{fig:simulated_psf}b, we simulate the optical system with the designed metasurface $r(p,q)$ and an achromatic refractive lens using DFlat~\cite{hazineh2022d}. The simulated PSF well approximates the desired optical model in Eq.~\ref{eq:optical-model-full}. The variation in energy distribution across these PSFs stems from deviations between the actual phase profile and its approximated counterpart. Light that bypasses modulation by the metasurface becomes residual light, which is focused solely by the achromatic refractive lens. In summary, the metasurface's dispersion induces a chromatic shift in the deflected PSF while also producing residual light focused by the refractive lens without any deflection.

\subsection{Image Formation}
Let $f(u,v,\lambda)$ represent the true hyperspectral datacube, 
the corresponding spatio-spectral irradiance profile $E(u,v)$ projected onto the full photosensor is:
\begin{align}
    \begin{split}
        &E(u,v,\lambda) =  f(u,v,\lambda) \odot \sum_i \alpha_i h(u-u_i(\lambda),v-v_i(\lambda)) \\
    \end{split}
\end{align}
where $\odot$ is the 2D convolution on $(u,v)$. 
The full image measured by the photosensor $I(u,v)$ is mathematically~\cite{brookshire2024metahdr}:
\begin{align}
\begin{split}
        I(u,v)=&G\cdot\text{Poisson}\left (t \cdot \int \eta(\lambda) E(u,v,\lambda)d\lambda\right) \\
        &+ \text{Gauss}(0,\sigma^2),
\end{split}
\end{align}
where $\eta(\lambda)$ is the photon efficiency at wavelength $\lambda$, $G$ is the gain, and $t$ is the exposure time, $\sigma$ is standard deviation of the read noise. In practice, we set the photon efficiency to be constant for all wavelengths. Similar to~\cite{brookshire2024metahdr}, we crop each sub-image centered at position $(u_i(\lambda_c),v_i(\lambda_c))$ from the full image, which we denote as $I_i(u,v)$, as shown in Fig.~\ref{fig:teaser}.

\subsection{Reconstruction} 
Various post-processing algorithms can be used to reconstruct H2 after the proposed image formation process. For instance, we follow the reconstruction procedure outlined in a recent work~\cite{shi2024split} that utilizes the deep network architecture that is a variant of the Deep Wiener Deconvolution Network (DWDN) \cite{dong2020deep}. The algorithm first hallucinates a feature cube $\mathcal{I}_i\in \mathbb{R}^{H\times W\times \Lambda}$ from each measurement $I_i\in\mathbb{R}^{H\times W\times 1}$, where $H\times W$ is the dimension of the image and $\Lambda$ is the dimension of the hyperspectral distribution. It then performs a Wiener deconvolution on all feature cubes: 
\begin{align}
    \tilde{f}(x,y,\lambda)=\mathcal{F}^{-1}\left\{\frac{\sum\limits_{i=1}^9 \overline{\mathcal{F}(h_i)}\mathcal{F}(\mathcal{I}_i)}{\sum\limits_{i=1}^9\overline{\mathcal{F}(h_i)}\mathcal{F}(h_i)}\right\},
\end{align}
where $\mathcal{F}$ and $\mathcal{F}^{-1}$ are Fourier and inverse Fourier transforms in the spatial dimension. The deconvolved features $\tilde{f}(x,y,\lambda)$ are further refined through a U-Net to produce the final hyperspectral datacube $\hat{f}(x,y,\lambda)$, each 2D slice being an HDR image. In Sec.~\ref{secsec:hyper-exp}, we also analyze other hyperspectral reconstruction algorithms on MetaH2 and show the adopted algorithm demonstrates the top reconstruction accuracy. 

\section{Experimental Results}
\begin{figure*}
    \vspace{-0.2in}
    \centering
    \includegraphics[width=\linewidth]{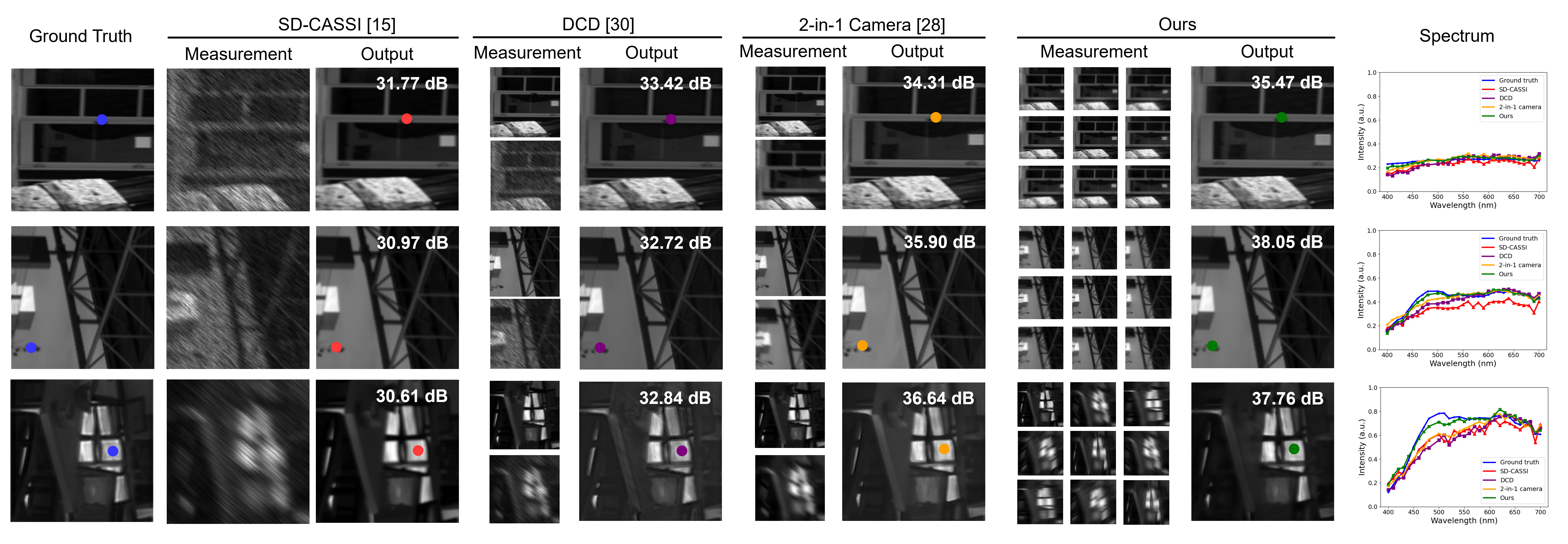}
    \vspace{-0.3in}
    \caption{Visual comparison between MetaH2 and previous snapshot hyperspectral imaging modalities on the ICVL dataset. We visualize the raw measurements, the output image at 550 nm, and the hyperspectral distribution at a pixel. Inset numbers show the PSNR, where MetaH2 constantly outperforms the rest. }
    \label{fig:comparison1}
\end{figure*}
\subsection{Snapshot Hyperspectral Imaging Comparison}
\label{secsec:hyper-exp}

First, we want to analyze how the proposed optical modality compares with previous snapshot hyperspectral imaging hardware. We simulate measurements from the single disperser CASSI (SD-CASSI) and its dual camera design (DCD) \cite{wang2015dual}, the split aperture 2-in-1 Camera~\cite{shi2024split}, and our system without the exposure bracketing, i.e., the sub-images sharing the same intensity. For SD-CASSI, a randomly generated binary mask serves as the coded aperture, while a grayscale camera is integrated to replicate the result of the DCD system. In the case of the 2-in-1 Camera, its optimized DOE features a linear dispersion in its PSF similar to our system. To make a uniform comparison, we set all methods to have uniform focal lengths, sensor dimensions, pixel pitch, and numerical aperture. Then, we reconstruct the hyperspectral datacube from the synthesized measurement of each method using four postprocessing algorithms~\cite{miao2019net,fu2021coded,dong2020deep,meng2020end} and compare the accuracy. The synthesized data are generated from two benchmark datasets, ICVL~\cite{arad2016sparse} and Harvard~\cite{chakrabarti2011statistics}, where we randomly select 9 and 50 images for testing and the remaining for training and validation, respectively.

Table~\ref{tab:comparison} lists the number of sub-images measured in a snapshot for each method. Considering a fixed photosensor dimension, MetaH2 trades the system's field of view (FoV) for more sub-images, resulting in the highest reconstruction accuracy universally under different metrics and on both benchmark datasets. Fig.~\ref{fig:comparison1} shows sample raw measurements of each system and the corresponding hyperspectral reconstruction. This analysis suggests MetaH2 is complementary to previous snapshot hyperspectral imaging solutions, as it measures more sub-images at the expense of FoV to generate a higher reconstruction accuracy.  

\begin{table*}[]
\vspace{0.1in}
\caption{Quantitative analysis of MetaH2 and previous snapshot hyperspectral imaging modalities. MetaH2 forms the most sub-images in a snapshot, leading to the highest hyperspectral reconstruction accuracy across all four postprocessing algorithms. The numbers in each cell are PSNR($\uparrow$) and SSIM($\uparrow$), respectively. }
\resizebox{1\textwidth}{!}{
\begin{tabular}{|c|c|cccc|cccc|}
\hline
\multirow{2}{*}{} & \multirow{2}{*}{\begin{tabular}[c]{@{}c@{}}\# sub-\\ images\end{tabular}} & \multicolumn{4}{c|}{ICVL} & \multicolumn{4}{c|}{Harvard} \\ \cline{3-10} 
 &  & \multicolumn{1}{c|}{$\lambda$-Net~\cite{miao2019net}} & \multicolumn{1}{c|}{DWDN~\cite{dong2020deep}} & \multicolumn{1}{c|}{DEIL~\cite{fu2021coded}} & TSA-Net~\cite{meng2020end} & \multicolumn{1}{c|}{$\lambda$-Net} & \multicolumn{1}{c|}{DWDN} & \multicolumn{1}{c|}{DEIL} & TSA-Net \\ \hline
SD-CASSI~\cite{wagadarikar2008single} & 1 & \multicolumn{1}{c|}{28.77 / 0.920} & \multicolumn{1}{c|}{30.29 / 0.939} & \multicolumn{1}{c|}{30.17 / 0.945} & 30.22 / 0.936 & \multicolumn{1}{c|}{28.84 / 0.851} & \multicolumn{1}{c|}{30.18 / 0.888} & \multicolumn{1}{c|}{29.76 / 0.893} & 28.78 / 0.853 \\ \hline
DCD~\cite{wang2015dual} & 2 & \multicolumn{1}{c|}{29.57 / 0.952} & \multicolumn{1}{c|}{30.73 / 0.950} & \multicolumn{1}{c|}{30.88 / 0.962} & 30.75 / 0.944 & \multicolumn{1}{c|}{28.99 / 0.881} & \multicolumn{1}{c|}{30.31 / 0.884} & \multicolumn{1}{c|}{31.33 / 0.911} & 29.14 / 0.876 \\ \hline
2-in-1 Camera~\cite{shi2024split} & 2 & \multicolumn{1}{c|}{29.62 / 0.957} & \multicolumn{1}{c|}{31.42 / 0.952} & \multicolumn{1}{c|}{31.51 / 0.957} & 31.18 / 0.960 & \multicolumn{1}{c|}{29.28 / 0.899} & \multicolumn{1}{c|}{30.56 / 0.905} & \multicolumn{1}{c|}{31.75 / 0.922} & 29.88 / 0.885 \\ \hline
Ours & 9 & \multicolumn{1}{c|}{\textbf{31.00/ 0.964}} & \multicolumn{1}{c|}{\textbf{31.62 / 0.968}} & \multicolumn{1}{c|}{\textbf{33.45 / 0.967}} & \textbf{31.23 / 0.965} & \multicolumn{1}{c|}{\textbf{30.45 / 0.902}} & \multicolumn{1}{c|}{\textbf{31.23 / 0.932}} & \multicolumn{1}{c|}{\textbf{32.20 / 0.937}} & \textbf{30.18 / 0.903} \\ \hline
\end{tabular}}
\label{tab:comparison}
\end{table*}

\vspace{-0.1in}
\begin{figure*}[!ht]
    \centering
    \includegraphics[width=\linewidth]{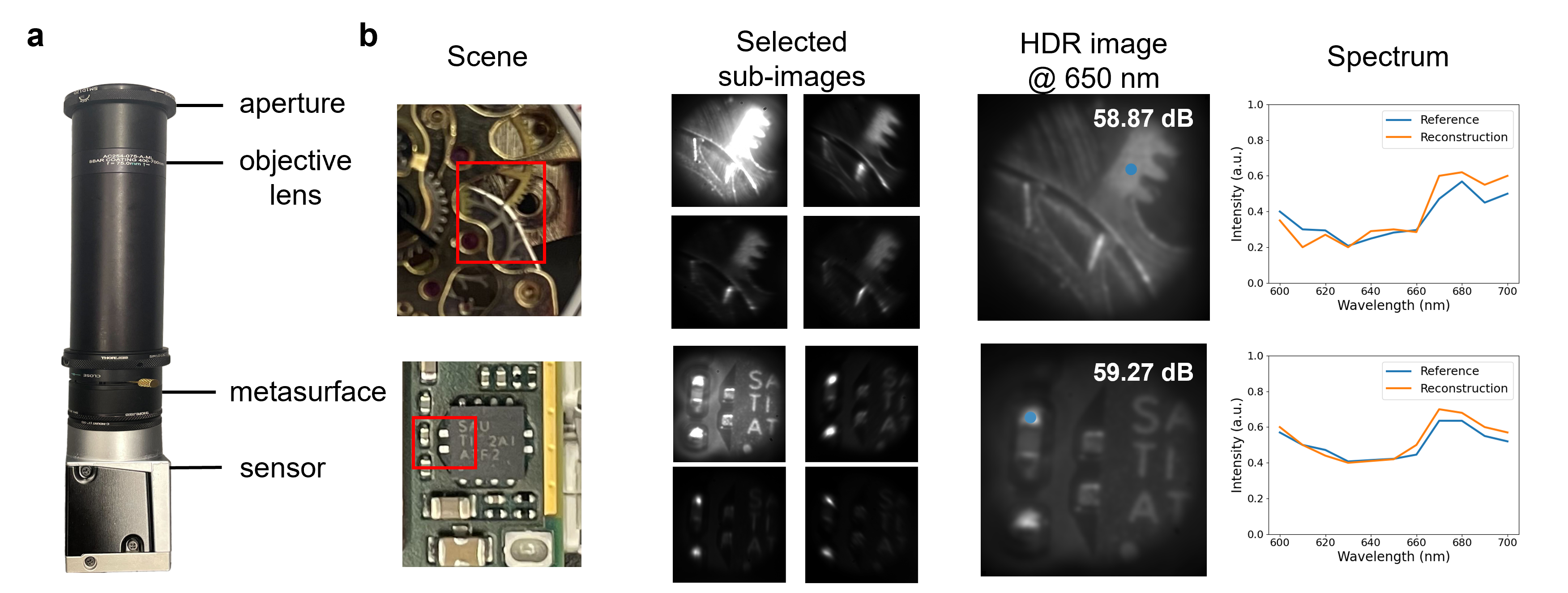}
    \caption{Prototype and real-world experiments. (a) The prototype camera. (b) Sample results showing how the camera captures HDR and hyperspectral data in a single snapshot, enabling measurement of dynamic objects (e.g., a moving gear) and highly reflective surfaces (e.g., circuit electronics). The inset number indicates the dynamic range of the reconstructed image, and the displayed spectrum shows the hyperspectral distribution at the pixel marked by the blue dot. }
    \label{fig:real_result}
\end{figure*}
\subsection{Prototyping and Real-world Experiments}
Fig.~\ref{fig:real_result} shows the working prototype of MetaH2 built upon Brookshire et al.~\cite{brookshire2024metahdr}. It consists of a 1-mm diameter metasurface that divides the light into nine regularly spaced positions on the photosensor.  Besides beamsplitting, the metasurface is designed to have a $20$~dpt optical power or $5$~cm focal length. Due to the small numerical aperture of the system, the designed metasurface is approximately equivalent to the achromatic lens and the beamsplitting metasurface in the MetaH2 optical design. The system consists of an objective lens to image close objects. 
We observed significant residual light in the central sub-image when the incident wavelength is shorter than 600 nm, possibly due to the phase discontinuity and fabrication defect of the metasurface. The major consequence of this residual light on our prototype is that the energy portion of the central image becomes disproportionately high compared to the designed value. Thus, we insert a bandpass filter to set the working bandwidth of the system to [600 nm, 700 nm]. Nonetheless, the bandwidth of MetaH2 is still much broader than many metasurface imaging systems, which typically have a 10~nm bandwidth~\cite{brookshire2024metahdr}. 

Fig.~\ref{fig:real_result} shows the HDR and hyperspectral reconstruction of sample scenes measured in a lab environment. The objects we image contain reflective areas that require HDR measurement to capture their complete intensity profiles. The prototype demonstrates measuring 60 dB dynamic range and a 10 nm spectral resolution using a monochrome photosensor.

\section{Conclusion}
In this paper, we introduce MetaH2, a metasurface-based camera designed for joint hyperspectral and HDR imaging. Our metasurface encodes both types of information into spatially multiplexed measurements, each characterized by distinct chromatic aberration and power level. Quantitative studies show that MetaH2 achieves universally higher hyperspectral imaging accuracy compared to other snapshot techniques at the cost of reduced FOV. We further build a working prototype demonstrating snapshot HDR and hyperspectral reconstruction on real-world scenes.

\vspace{0.2in}
\noindent \textbf{Acknowledgement.} The authors thank SNOChip Inc. for fabricating the metasurface.

\bibliographystyle{IEEEbib}
\footnotesize
\bibliography{main}

\end{document}